\documentclass[a4paper,twocolumn,11pt]{quantumarticle}
\pdfoutput=1
\usepackage[utf8]{inputenc}
\usepackage[english]{babel}
\usepackage[T1]{fontenc}
\usepackage[numbers]{natbib}
\usepackage{listings}

\usepackage{amsmath}
\usepackage{amssymb}
\usepackage{hyperref}
\usepackage[resetlabels]{multibib}

\begin{document}

\title{Semiquantum key distribution using initial states in only one basis without the classical user measuring}

\author{Xueying Liang}
\affiliation{School of Mathematics and Computational Science, Wuyi University, Jiangmen 529020, China}
%\orcid{0000-0002-2445-2701}
\author{Xiangfu Zou}
\email{xf.zou@hotmail.com}
%\homepage{http://quantum-journal.org}
%\orcid{0000-0003-0290-4698}
%\thanks{You can use the \texttt{\textbackslash{}email}, %\texttt{\textbackslash{}homepage},
%and \texttt{\textbackslash{}thanks} commands to add additional information for the preceding \texttt{\textbackslash{}author}. If applicable, this can also be used to indicate that a work has previously been published in conference proceedings.}
\affiliation{School of Mathematics and Computational Science, Wuyi University, Jiangmen 529020, China}
\author{Xin Wang}
\email{x.wang@cityu.edu.hk}
%\homepage{http://quantum-journal.org}
%\orcid{0000-0003-0290-4698}
%\thanks{You can use the \texttt{\textbackslash{}email}, %\texttt{\textbackslash{}homepage},
%and \texttt{\textbackslash{}thanks} commands to add additional information for the preceding \texttt{\textbackslash{}author}. If applicable, this can also be used to indicate that a work has previously been published in conference proceedings.}
\affiliation{City University of Hong Kong Shenzhen Research Institute, Shenzhen, Guangdong 518057, China}
\affiliation{Department of Physics, City University of Hong Kong, Tat Chee Avenue, Kowloon, Hong Kong SAR, China}
\author{Shenggen Zheng}
\affiliation{Peng Cheng Laboratory, Shenzhen, 518000, China}
%\orcid{0000-0003-1985-4623}
\author{Zhenbang Rong}
\affiliation{Faculty of Intelligent Manufacturing, Wuyi University, Jiangmen, 529020, China}
\author{Zhiming Huang}
\affiliation{School of Economics and Management, Wuyi University, Jiangmen, 529020, China}
%\orcid{0000-0003-1533-8015}
\author{Jianfeng Liu}
\affiliation{School of Mathematics and Computational Science, Wuyi University, Jiangmen 529020, China}
%\orcid{0000-0002-0335-9508}
\author{Ying Chen}
\affiliation{School of Mathematics and Computational Science, Wuyi University, Jiangmen 529020, China}
\author{Jianxiong Wu}
\affiliation{School of Mathematics and Computational Science, Wuyi University, Jiangmen 529020, China}
\maketitle

\begin{abstract}
From the perspective of resource theory, it is interesting to achieve the same quantum task using as few quantum resources as possible.
Semiquantum key distribution (SQKD), which allows a quantum user to share a confidential key with a classical user who prepares and operates qubits in only one basis, is an important example for studying this issue.
To further limit the quantum resources used by users, in this paper, we constructed the first SQKD protocol which restricts the quantum user to prepare quantum states in only one basis and removes the classical user's measurement capability.
Furthermore, we prove that the constructed protocol is unconditionally secure by deriving a key rate expression of the error rate in the asymptotic scenario.
The work of this paper provides inspiration for achieving quantum superiority with minimal quantum resources.
\end{abstract}
\section{Introduction}\label{sec1}
The information processing of quantum systems can accomplish some tasks that classical systems cannot, such as quantum cryptography~\cite{1983Conjugate}.
For instance, quantum key distribution (QKD), exploiting the quantum mechanical principle, can discover the disturbs caused by the observation of eavesdroppers~\cite{nielsen2010quantum}, while classical key distribution cannot.
Thus, if there is an eavesdropper listening in while Alice and Bob are sharing their key, the presence of the eavesdropper will be visible as a disturbance of the quantum channel used to establish the key.
QKD can implement a quantum cryptography task which is used to share a high security key between two legitimate users~\cite{bennett1984quantum,1983Conjugate}.
Following the idea in Ref.~\cite{1983Conjugate}, Bennett and Brassard~\cite{bennett1984quantum} proposed the first QKD protocol (BB84) which uses the unique properties of quantum mechanics, such as the quantum no-cloning theorem and the Heisenberg uncertainty principle to ensure the security of the key.
Furthermore, Bennett~\cite{B92} proposed a QKD protocol (B92) in which the initiator of the key sharing, Alice, only needs to prepare two nonorthogonal states.
Bru{\ss}~\cite{1998Six} proposed a six state protocol, which is safer against eavesdropping on single qubits than the one based on two conjugate bases.
After that, various QKD protocols based on discrete variables~\cite{2005Fast,branciard2008upper,cabello2000quantum,2021A} and continuous variables~\cite{grosshans2002continuous,leverrier2011continuous,zhou2022measurement} were proposed.

QKD can achieve unconditional security, while existing classical key distribution can generally achieve only computational security.
The security of a cryptographic protocol with unconditional security is independent of the algorithms and computing resources used by attackers~\cite{2001Unconditional,1999Unconditional,2000Simple}.
Contrastively, protocols that only have computational security will become increasingly insecure with the improvement of attackers' computing powers.
BB84 has been proven to be unconditionally secure~\cite{1999Unconditional,2000Simple}.
Lo and Chau~\cite{1999Unconditional} constructed a QKD protocol based on entanglement purification, being almost equivalent to BB84, and proved its unconditional security over arbitrarily long distances if there are fault-tolerant quantum computers.
Furthermore, Shor and Preskill~\cite{2000Simple} gave a entanglement-purification based  QKD protocol, using Calderbank-Shor-Steane codes, was shown to be unconditionally secure.
The proof removes the use of fault-tolerant quantum computers from the Lo-Chau proof~\cite{1999Unconditional}.
%This implies that BB84 is unconditionally secure over arbitrarily long distances.
More detailed discussions on the security of QKD can be found in Refs.~\cite{2001Unconditional,Biham2006Proof,Scarani2009security,Yin2016Security,Curty2019Simple,Xu2020Quantum}.
Note that, the one-time pad encryption is unconditionally secure when the length of the encryption key is not less than the length of the message~\cite{1949Communication}.
The difficulty of using the one-time pad encryption is how to obtain a sufficiently long secret key.
Fortunately, QKD can offer arbitrarily long encryption keys.
Accordingly, we can obtain unconditionally secure encryption methods by combining QKD technology with the one-time pad encryption.

Since observation in general disturbs the quantum system being observed, the third-party eavesdropping can be detected by the two parties involved in communication.
In classical key distribution, eavesdroppers cannot be detected by the users.
QKD protocols, such as BB84 and B92, require both Alice and Bob to prepare or operate non-orthogonal quantum states.
Currently, quantum resources are relatively expensive.
Simultaneously, quantum computing is not easy to implement, either.
Thereby, it is interesting to achieve the same quantum task using as few quantum resources as possible.
Generally, a quantum protocol having a significant advantage over all classical protocols, uses as few ``quantum'' as possible, which has aroused great interest among scholars.
In order to answer this question in quantum cryptography, Boyer et al.~\cite{boyer2007quantum} introduced the concept of semiquantum cryptography and constructed the first semiquantum key distribution (SQKD)  protocol (BKM2007), in which Bob is classical.
They also proved BKM2007's complete robustness.
Complete robustness means that there is a non-zero probability that legitimate participants find an error on the test bits if Eve gets non-zero information on the INFO string.
%Robustness of a protocol means that any adversarial attempt to learn some information on the key necessarily induces some observable disturbances.
In an SQKD protocol, the ``classical'' communication party is only allowed to perform some of four operations, i.e., (\romannumeral1) prepare qubits in the basis $\{\arrowvert 0\rangle,\,\arrowvert 1\rangle\}$, (\romannumeral2) measure qubits in the basis $\{\arrowvert 0\rangle,\,\arrowvert 1\rangle\}$, (\romannumeral3) reflect qubits without measuring, and (\romannumeral4) reorder qubits.
Lu an Cai~\cite{lu2008quantum} proposed an SQKD protocol with classical Alice and proved its complete robustness.
Boyer et al.~\cite{Boyer2009Semiquantum} proved the complete robustness of BKM2007 in a more general scenario that Alice sends qubits one by one, but sends the next one without waiting for a returning qubit.
They also proposed a randomization-based SQKD protocol and proved its complete robustness.
Zou et al.~\cite{2009Semiquantum} constructed five SQKD protocols in which Alice sends less than four quantum states, and proved that these protocols are completely robust.
After that, various SQKD protocols have been proposed.

Due to one user being classical, SQKD is not easy to give full play to the advantages of entanglement.
Nevertheless, some scholars have proposed some SQKD protocols using entangled states was proposed.
Wang et al.~\cite{wang2011semiquantum} proposed an SQKD protocol by using maximally entangled states, which can improve the qubit efficiency of the protocol.
Yu et al.~\cite{yu2014authenticated} proposed two SQKD protocols via Bell states, without using authenticated classical channels.
Zhou et al.~\cite{Zhou2019Multi} proposed an multi-party SQKD protocol based on four-particle cluster states, and proved its unconditional security.
Pan~\cite{pan2022semi} proposed a measure-resend SQKD protocol, only using three types of two-physical-qubit entangled states.

In general, single states is easier to be prepared than entangled states.
Some SQKD protocols, using single photon states based on one or more bases, were proposed.
For example, Krawec~\cite{Krawec2014Restricted} proposed a single-state SQKD protocol in which Alice just sends one quantum state to the classical Bob.
Wang et al.~\cite{wang2019efficient} proposed an SQKD protocol in which Alice sends two nonorthogonal states and provided the proof of its unconditional security.
Amer and Krawec~\cite{Amer2019Semiquantum} constructed an SQKD
protocol, in which Alice can work with two or three bases and proved its unconditional security.
Zhang et al.~\cite{zhang2020single} proposed a single-state SQKD protocol in which Alice just sends one qubit to the classical Bob and Bob just prepares one state in the preparation process, and proved its unconditional security.

In general, an SQKD protocol includes a classical user and a quantum user.
If a protocol only includes two users and both users are classical, it cannot share an secure key between the two classical users.
However, with the help of an untrusted quantum server, two classical users can share an unconditionally secure key.
Some mediated SQKD protocols~\cite{2014Mediated,2019Mediated,Krawec2019Multi}, using the untrusted quantum server, have been proposed.
Krawec~\cite{2014Mediated} proposed an mediated SQKD protocol, allowing two classical users to establish a secret key with the help of a fully quantum server, and proved its unconditional security.
Subsequently, a mediated SQKD protocol is proposed by Hwang et al.~\cite{2019Mediated}, allowing two classical participants to share a secret key with the help of an untrusted third party.
Furthermore, Krawec~\cite{Krawec2019Multi} constructed a multi-mediated SQKD protocol where two (or more) adversarial quantum servers are used, and proved its unconditional security.

Further restricting the ability of SQKD users to implement key distribution is an interesting problem.
Currently, some SQKD protocols~\cite{sun2013quantum,zou2015semiquantum,liu2018mediated} can remove the measurement capabilities of classical users.
Sun et al.~\cite{sun2013quantum} constructed two SQKD protocols that exempt classical Bob from measurement, and proved their complete robustness.
Zou~\cite{zou2015semiquantum} constructed an SQKD protocol, in which the classical Alice without measurement capability reorders qubits, and  proved its completely robust.
Liu and Hwang~\cite{liu2018mediated} proposed an SQKD protocol, allowing two ``classical'' participants without measurement capability to establish a shared secret key under an untrusted third party.

%Generally, there is one classical user and one quantum user in an SQKD protocol.
%SQKD restricts its classical users to prepare and operate quantum states on only one basis.
%Similarly, further restricting the ability of SQKD users to implement key distribution is an interesting problem, for example, restricting quantum users to prepare quantum states on only one basis, removing the measurement capabilities of classical users, and restricting all users to prepare and operate quantum states on only one basis but with the help of an untrusted quantum server.
Note that, the above proposed SQKD protocol restricts the quantum capabilities of one or both communication parties in various ways.
For example, restrict quantum users to prepare quantum state only in one basis~\cite{Krawec2014Restricted,zhang2020single}, remove the measurement capabilities of classical users~\cite{sun2013quantum,zou2015semiquantum,liu2018mediated}, or restrict all users to prepare and operate quantum states in only one basis but with the help of an untrusted quantum server~\cite{2014Mediated,2019Mediated,Krawec2019Multi}.
However, we find that there is no protocol with restricting the quantum user to prepare quantum states in only one basis and removing the measurement capability of the classical user. In this paper, we will construct an SQKD protocol with the two restrictions.

Complete robustness is a qualitative description between attackers obtaining useful information and legitimate users detecting errors.
It cannot provide the relationship between the amount of information obtained by attackers and the error rate detected by users.
There are many SQKD protocols having been proved to be completely robust ~\cite{boyer2007quantum,lu2008quantum,Boyer2009Semiquantum,2009Semiquantum,zou2015semiquantum}.
Note that, the final key rate is equal to the mutual information between legitimate users minus the amount of information obtained by the attacker.
Only when we know the relationship between the amount of information obtained by the attacker and the error rate detected by the users can we obtain the relationship between the final key rate and the error rate.
Unconditional security is described by the relationship between the final key rate and the detected error rate.
The characteristics of an SQKD protocol makes their security analysis more difficult than that of the corresponding full quantum counterpart.
SQKD depends on a two-way quantum communication channel between Alice and Bob.
This implies that the attacker, Eve, can perform two attacks.
Hence, the unconditional security proof of an SQKD protocol is more difficult than that of the corresponding full quantum QKD protocol.
From the above introduction, there are only the SQKD protocols~\cite{wang2019efficient,Amer2019Semiquantum,zhang2020single,Zhou2019Multi,2014Mediated,Krawec2019Multi} which have been shown to be unconditionally secure.
In this paper, we will show that the constructed protocol is unconditionally secure.

\textcolor{blue}
{
The main work of this article is to construct an unconditionally secure SQKD protocol that uses as few quantum resources as possible.
More specifically, the main contributions of this paper are as follows:
\begin{itemize}
  \item In the constructed SQKD protocol, the classical user Alice is not allowed to measure any quantum state. In other words, Alice is only allowed to prepare qubits in the basis $\{\arrowvert 0\rangle,\,\arrowvert 1\rangle\}$, reflect qubits without measuring, and reorder qubits.
  \item In the constructed SQKD protocol, the quantum user Bob is allowed to prepare initial qubits in only one basis.
      %, i.e., the basis $\{\arrowvert +\rangle,\,\arrowvert -\rangle\}$.
  \item The constructed SQKD protocol is showed to be unconditionally secure.
\end{itemize}
}

\section{Preliminaries}\label{sec2}
We set $\arrowvert +\rangle=\frac{1}{\sqrt{2}}(\arrowvert0\rangle\ +\arrowvert 1\rangle)$,  $\arrowvert -\rangle=\frac{1}{\sqrt{2}}(\arrowvert 0\rangle-\arrowvert 1\rangle)$, the $Z$ basis to be the basis
$\{\arrowvert 0\rangle,\,\arrowvert 1\rangle\}$, the $X$ basis to be the basis $\{\arrowvert +\rangle,\,\arrowvert -\rangle\}$, and $U^{*}$ to be the conjugate transposition of a unitary matrix $U$.

Suppose that $\{p_{1},p_{2},\cdots,p_{n}\}$ is a probability distribution, i.e., $p_{i}{\geq} 0 \ \textrm{and} \sum_{i}p_{i}{=}1$.
Then, the Shannon entropy
\begin{eqnarray}
H(p_{1},p_{2},{\cdots},p_{n}){=}{-}\sum_{i}p_{i}\log p_{i}.
\end{eqnarray}
In this paper, logarithms indicated by ``$\log$'' are taken to base two.
If $\rho$ is a density operator acting on the finite Hilbert space $\mathcal{H}$, %$S(\rho)$ is
its von Neumann entropy
\begin{eqnarray}
S(\rho)=-\sum_{i}\lambda_{i}\log\lambda_{i},
\end{eqnarray}
where $\lambda_{i}$ is the eigenvalues of $\rho$.
Let $\rho_{BA}$ be a density operator acting on the space $\mathcal{H}_{A}\otimes \mathcal{H}_{B}$. $S(AB)$ ($S(A)$) is denoted the von Neumann entropy of $\rho_{AB}$
 ($\rho_{A}=\textrm{tr}_{B}\rho_{AB}$).
 We will write $S(A\arrowvert B)$ to be the von Neumann entropy of $A$ conditional on knowing $B$, i.e., $S(A\arrowvert B)=S(AB)-S(B)=S(\rho_{AB})-S(\textrm{tr}_{A}(\rho_{AB}))$.

Let $N$ be the number of key bits of Alice and Bob before information reconciliation and privacy amplification in the protocol but after the error rate check which is called the raw key of Alice and Bob.
Let $\ell(N)\leq N$ denote the key size that Alice and Bob may distill in the future. Then, by Refs.~\cite{devetak2005distillation,renner2005information,krawec2015security}, we know the key rate
\begin{eqnarray}\label{eqr}
r=\lim_{N\rightarrow\infty}\frac{\ell(N)}{N}\geq \inf(S(A\arrowvert E)-H(A\arrowvert B)),
\end{eqnarray}
\newline
or
\begin{eqnarray}\label{eqr}
r=\lim_{N\rightarrow\infty}\frac{\ell(N)}{N}\geq \inf(S(B\arrowvert E)-H(B\arrowvert A)),
\end{eqnarray}
where the minimum is calculated under all collective attacks allowed by the error detection of legitimate parties.
\section{The SQKD protocol}\label{sec3}
Although there are SQKD protocols in which the quantum user prepares quantum states in only one basis or the classical user does not need to measure, there is no protocol in which the quantum user prepares quantum states in only one basis and the classical user does not need to measure.
In this section, we construct this type of SQKD protocol.
More specifically, we construct an SQKD protocol in which Alice is a limited ``classical'' user without measurement capability and a quantum user, Bob, may prepare qubits in only the $X$ basis but measure qubits in any basis of his choice.
The SQKD protocol consists of the following procedure.

Step~1. Bob prepares and sends to Alice $N=4n(1+\delta)$ qubits, each one chosen randomly from $\{\arrowvert +\rangle, \ \arrowvert -\rangle\}$.
The sequence of these qubits is called $P_B$.

Step~2. After receiving all qubits from Bob, Alice chooses $M~(M\geq N)$ qubits from $\{\arrowvert 0\rangle,\arrowvert 1\rangle\}$ at random.
The sequence of these qubits is called $P_A$.
Then, she will reorder randomly all qubits in $P_A$ and $P_B$ and send the sequence of the first $2N$ qubits, $P_{AB}$, to Bob.
The qubits sent by Bob in $P_{AB}$ is called CTRL qubits and that inserted by Alice SIFT qubits.

Step~3. Bob chooses randomly the $Z$ basis or the $X$ basis to measure each qubit.

Step~4. Alice informs Bob, utilizing the classical channel, the order of the qubit sequence $P_{AB}$ and Bob publishes the positions which he measured in the $Z$ basis.
The results, which Bob measured the SIFT qubits in the $Z$ basis, are called SIFT-$Z$ bits.
Contrastively, the results, which Bob measured the CTRL qubits in the $X$ basis, are called CTRL-$X$ bits.
If the number of SIFT-$Z$ bits is not less than $2n$, Alice and Bob go to Step 1.

Step~5. Bob checks the error rate on the CTRL qubits. When Bob measured with the $X$ basis, the measurement result state needs to be same as he sent originally.
Otherwise, it is seen as an error.
If the error rate exceeds the predefined threshold $T_{X}$, Alice and Bob terminate the protocol.

Step~6. Bob chooses randomly $n$ SIFT-$Z$ bits as TEST bits and announces their positions and values by the classical channel.
Bob's measurement result states must be same as those prepared by Alice. Alice checks the error rate of the TEST bits. If the error rate exceeds the predefined threshold $T_{Z}$, Alice and Bob terminate the protocol.

Step~7. Alice and Bob select the first $n$ remaining SIFT-$Z$ bits as raw key.

Step~8. Alice and Bob run error correcting and privacy amplification on the raw key to distill the final key.
\section{Security proof}\label{sec4}
In this section, we show that the constructed SQKD protocol can resist collective attacks, i.e., Eve performs the same attack operation on the transit particle and the corresponding ancilla in each iteration, and sends the transit particle.
Note that, Alice and Bob use the SIFT-$Z$ bits as their raw key.
For each SIFT-$Z$ bit, Alice's key bit is $i$ if she prepares $\arrowvert i\rangle$ in Step 2, and Bob's key bit is $j$ if his measurement result state is $\arrowvert j\rangle$ in Step 3.

Let $\mathcal{H}_{T}$ be the two-dimensional Hilbert space modeling the transit qubit (the transit space).
 %and let $\mathcal{H}_{E_1}\otimes\mathcal{H}_{E_2}$ be Eve's ancilla space for one iteration of protocol.
We use the particle, $E_1$ ($E_2$), to represent Eve's ancilla particle which assists to attack the qubit traveling from Bob to Alice (from Alice to Bob) for one iteration.
The state of $E_1$ ($E_2$) is in the space $\mathcal{H}_{E_1}$ ($\mathcal{H}_{E_2}$).
%while $\mathcal{H}_{E_2}$ be Eve's ancilla space which is generated by the particles used to attack the the qubit returning from Alice to Bob for one iteration of protocol.
Let $U_{E}$ be an attack operator which acts on $\mathcal{H}_{T}\otimes \mathcal{H}_{E_1}$, and $U_{F}$ an attack operator which acts on $\mathcal{H}_{T}\otimes \mathcal{H}_{E_2}$.
In order to estimate the protocol's key rate, we first construct the density operator describing the joint state when Alice sends a state and Bob measures it in the $Z$ basis.
This case can contribute a bit to the raw key.

In this event, Alice prepares a qubit in the $Z$ basis, each of $\arrowvert 0\rangle$ and $\arrowvert 1\rangle$ chosen with probability $\frac{1}{2}$.
Then, the state Eve receives can be seem as
\begin{equation}
 \rho_{0} = \frac{1}{2} \arrowvert0\rangle \langle0\arrowvert_{T} + \frac{1}{2} \arrowvert1\rangle \langle1\arrowvert_{T}.
\end{equation}

We may assume Eve's ancilla is cleared to some $\arrowvert 0\rangle_{E_{2}}\in \mathcal{H}_{E_2}$. Eve then attacks with $U_{F}$, an operator which acts on basis states as
\begin{eqnarray}\label{eq1}
 &U_{F}\arrowvert0,0\rangle_{E_{2}T}=\arrowvert e_{0,0}^{A},0\rangle_{E_{2}T}+\arrowvert e_{0,1}^{A},1\rangle_{E_{2}T},\\
 &U_{F}\arrowvert0,1\rangle_{E_{2}T}=\arrowvert e_{1,0}^{A},0\rangle_{E_{2}T}+\arrowvert e_{1,1}^{A},1\rangle_{E_{2}T},
\end{eqnarray}
such that
\begin{equation}
 \langle e_{0,0}^{A}\arrowvert e_{0,0}^{A}\rangle {+} \langle e_{0,1}^{A}\arrowvert e_{0,1}^{A}\rangle {=} 1\!,
\end{equation}
\begin{equation}
\langle e_{1,0}^{A}\arrowvert e_{1,0}^{A}\rangle {+} \langle e_{1,1}^{A}\arrowvert  e_{1,1}^{A}\rangle {=} 1\!,
\end{equation}
\begin{equation}
 \langle e_{0,0}^{A}\arrowvert e_{1,0}^{A}\rangle+\langle e_{0,1}^{A}\arrowvert e_{1,1}^{A}\rangle=0.
\end{equation}

After this operation, Eve passes the transit qubit to Bob. Bob performs a $Z$ basis measurement. Then, the system collapses into
\begin{eqnarray}
%\begin{aligned}
 \rho_{BE_{2}T}{=}\frac{1}{2}\arrowvert0\rangle\langle 0\arrowvert_{B}{\otimes}(\arrowvert e_{0,0}^{A},0\rangle\langle e_{0,0}^{A},0\arrowvert \nonumber\\
 {+}\arrowvert e_{1,0}^{A},0\rangle\langle e_{1,0}^{A},0\arrowvert)_{E_{2}T}\nonumber
 \\
 {+}\frac{1}{2}\arrowvert1\rangle\langle1\arrowvert_{B}{\otimes}(\arrowvert e_{0,1}^{A},1\rangle\langle e_{0,1}^{A},1\arrowvert\nonumber\\
 {+}\arrowvert e_{1,1}^{A},1\rangle\langle e_{1,1}^{A},1\arrowvert)_{E_{2}T}.
\end{eqnarray}
Tracing out the system $T$ from $\rho_{BE_{2}T}$, we have
\begin{eqnarray}\label{eq2}
%\begin{aligned}
 \rho_{BE_{2}}{=}\frac{1}{2}\arrowvert0\rangle\langle 0\arrowvert_{B}{\otimes}(\arrowvert e_{0,0}^{A}\rangle\langle e_{0,0}^{A}\arrowvert{+}\arrowvert e_{1,0}^{A}\rangle\langle e_{1,0}^{A}\arrowvert)_{E_{2}}\nonumber\\
 +\frac{1}{2}\arrowvert1\rangle\langle1\arrowvert_{B}{\otimes}(\arrowvert e_{0,1}^{A}\rangle\langle e_{0,1}^{A}\arrowvert{+}\arrowvert e_{1,1}^{A}\rangle\langle e_{1,1}^{A}\arrowvert)_{E_{2}}\!.
 %\end{aligned}
\end{eqnarray}

It's important to observe that Alice and Bob may estimate the SIFT bits noise, during the parameter estimation stage. In particular, they estimate the quantity $p_{i,j}^{A}$ which we use to denote the probability that, if Alice sends $\arrowvert i\rangle$, then Bob's measurement result state is $\arrowvert j\rangle$, $ i, j\in\{0,1\}$. For example, if there is no noise in the SIFT bits, it should hold that $p_{0,0}^{A}=p_{1,1}^{A}=1$ and $p_{0,1}^{A}=p_{1,0}^{A}=0$. These parameters can be used to estimate the value $\langle e_{a,b}^{A}\arrowvert e_{a,b}^{A}\rangle$.

For example, to estimate $p_{0,1}^{A}$, consider the case that Alice first send $\arrowvert0\rangle$. After Eve 's attack, the state evolves to $\arrowvert e_{0,0}^{A},0\rangle+\arrowvert e_{0,1}^{A},1\rangle$ and the probability of Bob's measurement is $\arrowvert1\rangle$ is $\langle e_{0,1}^{A}\arrowvert e_{0,1}^{A}\rangle$, i.e.~$\langle e_{0,1}^{A}\arrowvert e_{0,1}^{A}\rangle=p_{0,1}^{A}$. Similarly, we have $\langle e_{0,0}^{A}\arrowvert e_{0,0}^{A}\rangle=p_{0,0}^{A}$, $\langle e_{1,0}^{A}\arrowvert e_{1,0}^{A}\rangle=p_{1,0}^{A}$ and $\langle e_{1,1}^{A}\arrowvert e_{1,1}^{A}\rangle=p_{1,1}^{A}$.
%\begin{eqnarray}
% \langle e_{0,0}^{A}\arrowvert e_{0,0}^{A}\rangle{=}p_{0,0}^{A}\!,\!\langle e_{1,0}^{A}\arrowvert e_{1,0}^{A}\rangle{=}p_{1,0}^{A}\!,\!\langle e_{1,1}^{A}\arrowvert e_{1,1}^{A}\rangle{=}p_{1,1}^{A}\!.~~~~~
%\end{eqnarray}

To estimate the key rate, according to Eq.~(\ref{eqr}), we need to estimate $S(B\arrowvert E)$ and $H(B\arrowvert A)$.
Now, we compute $H(B \arrowvert A)=H(A,B)-H(A)$.
We use $p(0)$ to denote the probability that Alice's raw key bit is zero.
Then,
\begin{equation}
p(0)=\frac{1}{2}(p_{0,0}^{A}+p_{0,1}^{A})=\frac{1}{2}.
\end{equation}
Thus,
\begin{equation}
H(A)=H(p(0),1-p(0))=1.
\end{equation}
We use $p(a,b)$ to denote the probability that Bob's raw key bit is $b$ while Alice's is $a$. These values are
\begin{eqnarray}
p(0,0)=\frac{1}{2}p_{0,0}^{A},~~~~~p(0,1)=\frac{1}{2}p_{0,1}^{A},\\ %\nonumber\\
p(1,0)=\frac{1}{2}p_{1,0}^{A},~~~~~p(1,1)=\frac{1}{2}p_{1,1}^{A}.
\end{eqnarray}
Then, we have
\begin{eqnarray}
H(A,B){=}H(p(0,0),p(0,1),p(1,0),p(1,1)) .
\end{eqnarray}
Further,
\begin{eqnarray}
H(B \arrowvert A){=}H(p(0,\!0),p(0,\!1),p(1,\!0),p(1,\!1)){-}1.
\end{eqnarray}

Now, we just need to estimate $S(B\arrowvert E)$.

\subsection{$S(B\arrowvert E)$ analysis}\label{subsec1}
Due to know incompletely the states of Eve's the ancilla particles, Alice and Bob can only use their own particle states to estimate $S(B\arrowvert E)$.
We modify the technique in Ref.~\cite{krawec2015security} suitably for our purpose.

We introduce a new system $H_{C}$ into Eq. (\ref{eq2}). Due to the strong sub additivity of von Neumann entropy,
 \begin{equation}
 S(B\arrowvert E)\geq S(B\arrowvert EC).
\end{equation}
Let $H_{C}$ be the two dimensional space spanned by $\{\arrowvert C\rangle,\arrowvert W\rangle\}$.
We will use the state $\arrowvert C\rangle\langle C\arrowvert$ to stand for the event that Alice and Bob's raw bits agree (that is, the state Bob measures is same as the state Alice sends).
Similarly for the state $\arrowvert W\rangle\langle W\arrowvert$ where Alice's and Bob's raw key bits do not agree.
After the system $H_{C}$ is added, the joint system is
\begin{eqnarray}
&\rho_{BE_{2}C}& \!\!\!\! =\frac{1}{2}\arrowvert0\rangle\langle0\arrowvert_{B}\otimes(\arrowvert e_{0,0}^{A}\rangle\langle e_{0,0}^{A}\arrowvert_{E_{2}}\otimes \arrowvert C\rangle\langle C\arrowvert_{C} \nonumber\\
&& +\arrowvert e_{1,0}^{A}\rangle\langle e_{1,0}^{A}\arrowvert_{E_{2}} \otimes \arrowvert W\rangle\langle W\arrowvert_{C}) \nonumber\\
&& +\frac{1}{2}\arrowvert1\rangle\langle1\arrowvert_{B}\otimes(\arrowvert e_{0,1}^{A}\rangle\langle e_{0,1}^{A}\arrowvert_{E_{2}} \otimes\arrowvert W\rangle\langle W\arrowvert_{C}~\nonumber\\
&& +\arrowvert e_{1,1}^{A}\rangle\langle e_{1,1}^{A}\arrowvert_{E_{2}} \otimes \arrowvert C\rangle\langle C\arrowvert_{C})\nonumber\\
&&\!\!\!\! =YDY^{*},
\end{eqnarray}
where $Y=(\arrowvert0,e_{0,0},C\rangle,\arrowvert1,e_{0,1},W\rangle,\linebreak[3]\arrowvert1,e_{1,1},C\rangle, \linebreak[3]\arrowvert0,e_{1,0},W\rangle)$.
%\begin{eqnarray}
%Y{=}(\arrowvert0\!,e_{0,0}\!,C\rangle\!,\arrowvert1\!,e_{0,1}\!,W\rangle\!,\arrowvert\!1,e_{1,1}\!,C\rangle\!, \arrowvert0\!,e_{1,0}\!,W\rangle\!)
%\end{eqnarray}
 and
\begin{eqnarray}
D =
 \left(
   \begin{array}{cccc}
     \frac{1}{2} p^{A}_{0,0}& 0 & 0 & 0 \\
     0 & \frac{1}{2} p^{A}_{0,1} & 0 & 0 \\
     0 & 0 & \frac{1}{2} p^{A}_{1,1} & 0 \\
     0 & 0 &  & \frac{1}{2} p^{A}_{1,0} \\
   \end{array}
 \right).
\end{eqnarray}
If $\arrowvert e^{A}_{i,j}\rangle$ is a non-zero vector, $\arrowvert e_{i,j}\rangle{=}\frac{\arrowvert e^{A}_{i,j}\rangle}{\langle e^{A}_{i,j}\arrowvert e^{A}_{i,j}\rangle}$; Otherwise, $\arrowvert e_{i,j}\rangle{=}\boldsymbol0$.
Thus,
%Choosing the basis $\{\arrowvert0,e_{0,0},C\rangle,\arrowvert0,e_{0,1},W\rangle,\arrowvert1,e_{1,0},C\rangle,$ $\arrowvert1,e_{1,0},W\rangle\}$, the matrix representation of $\rho_{BCE_{2}}$ is the diagonal matrix $D$.
\begin{eqnarray}
S(BE_{2}C)=H(\frac{1}{2} p^{A}_{0,0},\frac{1}{2} p^{A}_{0,1},\frac{1}{2} p^{A}_{1,1},\frac{1}{2} p^{A}_{1,0}).
\end{eqnarray}

This is a quantity that Alice and Bob may compute after the parameter estimation stage.

Tracing out $B$ from $\rho_{BEC}$, we have
\begin{equation}
 \rho_{E_{2}C}=\frac{1}{2}\sigma_{1}\otimes\arrowvert C\rangle\langle C\arrowvert+ \frac{1}{2}\sigma_{2}\otimes\arrowvert W\rangle\langle W\arrowvert,
\end{equation}
where $\sigma_{1}=\arrowvert e_{0,0}^{A}\rangle\langle e_{0,0}^{A}\arrowvert+\arrowvert e_{1,1}^{A}\rangle\langle e_{1,1}^{A}\arrowvert$ and $\sigma_{2}=\arrowvert e_{0,1}^{A}\rangle\langle e_{0,1}^{A}\arrowvert+\arrowvert e_{1,0}^{A}\rangle\langle e_{1,0}^{A}\arrowvert$.

Assume that $\textrm{tr}(\sigma_{1}),\textrm{tr}(\sigma_{2})>0$.
Note that, $\textrm{tr}\sigma_{1}=0$ and $\textrm{tr}\sigma_{2}=0$ are equivalent to $ p^{A}_{0,0}+ p^{A}_{1,1}=0$ and $ p^{A}_{0,1}+ p^{A}_{1,0}=0$, respectively.
Furthermore, $ p^{A}_{0,0}= p^{A}_{1,1}=0$ if and only if $\arrowvert e_{0,0}^{A}\rangle=\arrowvert e_{1,1}^{A}\rangle=\boldsymbol0$; $ p^{A}_{0,1}= p^{A}_{1,0}=0$ if and only if $\arrowvert e_{0,1}^{A}\rangle=\arrowvert e_{1,0}^{A}\rangle=\boldsymbol0$.
Thus, if $\textrm{tr}\sigma_{1}=0$ ($\textrm{tr}\sigma_{2}=0$), $\sigma_{1}$ ($\sigma_{2}$) could simply be removed from the description of $\rho_{E_{2}C}$ above. Moreover, any item related to $\sigma_{1}$ ($\sigma_{2}$) is removed from the subsequent computation of $S(E_{2}C)$.
Let $\widetilde{\sigma}_{j}=\frac{\sigma_{j}}{t_{j}}$, $t_{j}=\textrm{tr}(\sigma_{j})$.
Then,
\begin{equation}\label{re2c}
\rho_{E_{2}C}=\frac{1}{2}t_{1}\widetilde{\sigma}_{1}\otimes\arrowvert C\rangle\langle C\arrowvert+\frac{1}{2}t_{2}\widetilde{\sigma}_{2}\otimes\arrowvert W\rangle\langle W\arrowvert.
\end{equation}

Since $\textrm{tr}(\rho_{E_2C})=\textrm{tr}(\widetilde{\sigma}_{1})=\textrm{tr}(\widetilde{\sigma}_{2})=1$, we get $\frac{1}{2}t_{1}+\frac{1}{2}t_{2}=1$. Applying Lemma~1 in Ref.~\cite{krawec2015security} to Eq.~(\ref{re2c}),
\begin{eqnarray}
S(E_{2}C)&&\!\!\!\!\!\!\!\!\!\!{=}S(\rho_{E_{2}C})\nonumber\\
&&\!\!\!\!\!\!\!\!\!\!{=}H(\frac{1}{2}t_{1},\frac{1}{2}t_{2}) {+} \frac{1}{2}t_{1}S(\widetilde{\sigma}_{1}) {+} \frac{1}{2}t_{2}S(\widetilde{\sigma}_{2})\!.~~~~~~
\end{eqnarray}

Since $\widetilde{\sigma}_{2}$ being a density operator of a two-dimensional system, $S(\widetilde{\sigma}_{2})\leq1$. Note that, $t_{1}=\textrm{tr}(\sigma_{1})= p^{A}_{0,0}+ p^{A}_{1,1}=p_{0,0}^{A}+p_{1,1}^{A}$ and $t_{2}=\textrm{tr}(\sigma_{2})= p^{A}_{0,1}+ p^{A}_{1,0}=p_{0,1}^{A}+p_{1,0}^{A}$. Thus,
 \begin{eqnarray}
S(E_{2}C){\leq} H(\frac{1}{2}(p_{0,0}^{A}{+}p_{1,1}^{A}),\frac{1}{2}(p_{0,1}^{A}{+}p_{1,0}^{A}))\nonumber\\
{+}\frac{1}{2}(p_{0,1}^{A}{+}p_{1,0}^{A}){+}\frac{1}{2}(p_{0,0}^{A}{+}p_{1,1}^{A})S(\widetilde{\sigma}_{1}).
\end{eqnarray}

In order to estimate $S(\widetilde{\sigma}_{1})$, we first calculate the eigenvalues of  $\sigma_{1}$. Let $\arrowvert e_{0,0}^{A}\rangle=\sqrt{p_{0,0}^{A}}\arrowvert \zeta\rangle$ and $\arrowvert e_{1,1}^{A}\rangle=\alpha\arrowvert \zeta\rangle+\beta\arrowvert\xi\rangle$, where $\langle \zeta\arrowvert \zeta\rangle=\langle\xi\arrowvert\xi\rangle=1$, $\langle \zeta\arrowvert\xi\rangle=0$, and $\alpha,\beta\in \mathcal{C}$. In the basis $\{\arrowvert \zeta\rangle,\arrowvert\xi\rangle\}$,
\begin{equation}
\sigma_{1}=\left(
                  \begin{array}{cc}
                    p_{0,0}^{A} +\arrowvert\alpha\arrowvert^{2}& \alpha\beta^{*} \\
                    \alpha^{*}\beta & \arrowvert\beta\arrowvert^{2} \\
                  \end{array}
                \right).
\end{equation}

By calculating, the eigenvalues of $\sigma_{1}$ are
\begin{eqnarray}
\lambda_{\pm}{=}\frac{p_{0,0}^{A}{+}p_{1,1}^{A}{\pm}\sqrt{(p_{0,0}^{A}{-}p_{1,1}^{A})^{2}{+}4\arrowvert\langle e_{0,0}^{A}\arrowvert e_{1,1}^{A}\rangle\arrowvert^{2}}}{2}\!.~
\end{eqnarray}

Since $\widetilde{\sigma}_{1}=\frac{\sigma_{1}}{p_{0,0}^{A}+p_{1,1}^{A}}$, the eigenvalues of $\widetilde{\sigma}_{1}$ are
\begin{equation}\label{lambda}
\widetilde{\lambda}_{\pm}=\frac{1}{2}\pm\frac{\sqrt{(p_{0,0}^{A}-p_{1,1}^{A})^{2}+4\arrowvert\langle e_{0,0}^{A}\arrowvert e_{1,1}^{A}\rangle\arrowvert^{2}}}{2(p_{0,0}^{A}+p_{1,1}^{A})}.
\end{equation}
Thus,
%\begin{eqnarray}\label{sigma1}
%S(\widetilde{\sigma}_{1})\!&=&\!-\widetilde{\lambda}_{+}\log\widetilde{\lambda}_{+} {-} \widetilde{\lambda}_{-}\log\widetilde{\lambda}_{-}\nonumber\\
%\!&=&\!h(\widetilde{\lambda}_{+})\nonumber\\
%\!&=&\!h\!\left(\frac{1}{2}+\frac{\sqrt{(p_{0,0}^{A}-p_{1,1}^{A})^{2}+4\arrowvert\langle e_{0,0}^{A}\arrowvert e_{1,1}^{A}\rangle\arrowvert^{2}}}{2(p_{0,0}^{A}+p_{1,1}^{A})}\right)\!.~~~~~
%\end{eqnarray}
\begin{eqnarray}\label{sigma1}
S(\widetilde{\sigma}_{1})&&\!\!\!\!\!\!\!\!\!\!{=}{-}\widetilde{\lambda}_{{+}}\log\widetilde{\lambda}_{{+}} {{-}} \widetilde{\lambda}_{{-}}\log\widetilde{\lambda}_{{-}}\nonumber\\
&&\!\!\!\!\!\!\!\!\!\!{=}h(\widetilde{\lambda}_{{+}})\nonumber\\
&&\!\!\!\!\!\!\!\!\!\!{=}h\!\!\left(\!\frac{1}{2}{+}\frac{\sqrt{(p_{0,0}^{A}{-}p_{1,1}^{A})^{2}{+}4\arrowvert\langle e_{0,0}^{A}\arrowvert e_{1,1}^{A}\rangle\arrowvert^{2}}}{2p_{0,0}^{A}{+}2p_{1,1}^{A}}\right)\!\!\!.~~~~~
\end{eqnarray}
By Eq.~(\ref{sigma1}), there is only $\arrowvert\langle e_{0,0}^{A} \arrowvert e_{1,1}^{A}\rangle\arrowvert^{2}$ unknown in the final expression of $S(\widetilde{\sigma}_{1})$.
Thus, we only need to estimate $\arrowvert\langle e_{0,0}^{A}\arrowvert e_{1,1}^{A}\rangle\arrowvert^{2}$.

We can determine a lower bound of $\arrowvert\langle e_{0,0}^{A}\arrowvert e_{1,1}^{A}\rangle\arrowvert^{2}$ by the noise in CRTL bits.
Note that,
Bob prepares both $\arrowvert+\rangle$ and $\arrowvert-\rangle$ with probability $\frac{1}{2}$.
The state Eve receives in the forward quantum channel can be seem as
\begin{equation}
\rho_{1}= \frac{1}{2} \arrowvert+\rangle \langle+\arrowvert_{T} + \frac{1}{2} \arrowvert-\rangle \langle-\arrowvert_{T}.
\end{equation}

Without loss of generality, We may assume that Eve's ancilla is cleared to $\arrowvert0\rangle_{E_{1}}\in H_{E_{1}}$. Then, Eve attacks with $U_{E}$ defined as
\begin{eqnarray}\label{eq4}
 U_{E}\arrowvert+,0\rangle_{TE_{1}}=\arrowvert+,e_{+,+}^{B}\rangle_{TE_{1}}+\arrowvert-,e_{+,-}^{B}\rangle_{TE_{1}},~
\end{eqnarray}
\begin{eqnarray}
 U_{E}\arrowvert-,0\rangle_{TE_{1}}=\arrowvert+,e_{-,+}^{B}\rangle_{TE_{1}}+\arrowvert-,e_{-,-}^{B}\rangle_{TE_{1}},~
\end{eqnarray}
with
\begin{eqnarray}
 \langle e_{+,+}^{B}\arrowvert e_{+,+}^{B}\rangle+\langle e_{+,+}^{B}\arrowvert e_{+,-}^{B}\rangle &=&1,\\
 \langle e_{-,+}^{B}\arrowvert e_{-,+}^{B}\rangle+\langle e_{-,-}^{B}\arrowvert  e_{-,-}^{B}\rangle &=&1,\\
 \langle e_{+,+}^{B}\arrowvert e_{-,+}^{B}\rangle+\langle e_{+,-}^{B}\arrowvert e_{-,-}^{B}\rangle &=&0.
\end{eqnarray}

For the convenience of the following discussion, let $p_{+,+}=\langle e_{+,+}^{B}\arrowvert e_{+,+}^{B}\rangle$, $p_{+,-}=\langle e_{+,-}^{B}\arrowvert e_{+,-}^{B}\rangle$, $p_{-,+}=\langle e_{-,+}^{B}\arrowvert e_{-,+}^{B}\rangle$, $p_{-,-}=\langle e_{-,-}^{B}\arrowvert e_{-,-}^{B}\rangle$.

 After performing the attack $U_{E}$, Eve passes the transit qubit to Alice.
 After receiving all qubits, Alice prepares some new qubits, reorders the prepared qubits and the received qubits, and sends the first $2N$ qubits to Bob.
 Note that, Eve can apply her second attack $U_{F}$ when Alice sends the qubits to Bob.

Let $\arrowvert b\rangle_{i}$ represent the $i$-th particle sent by Bob.
 Correspondingly, $\arrowvert e\rangle_{i}$ represents the ancilla when Eve attacks $\arrowvert b\rangle_{i}$.
 For convenience, the $i$-th ancilla that Eve uses to execute the attack $U_{E}$ is denoted by $\arrowvert e\rangle_{i}$.
 Note that, the probability that $\arrowvert e\rangle_{i}$ still matches $\arrowvert b\rangle_{i}$ after Alice's reordering operation is $\frac{1}{N+M}$ ($M\geq N$).
 When $N$ is large enough, the probability is close to zero. Thus, we can assume that Eve's second attack on $\arrowvert b\rangle_{i}$ is independent of $\arrowvert e\rangle_{i}$, i.e., using a new ancilla.

In the event, Bob sends the state $\arrowvert a\rangle$ and the state returns to him, Eve can attack $\arrowvert a\rangle$ twice.
Let $V=(U_{F}\otimes I_{E_{1}})(I_{E_{2}}\otimes U_{E})$.
Using Eqs. (\ref{eq1}) and (\ref{eq4}), $V$'s action on basis states $\arrowvert +\rangle,\arrowvert -\rangle\in H_{T}$ can be described as
\begin{eqnarray}
V&&\hskip -7mm
(\arrowvert0\rangle_{E_{2}}\arrowvert+\rangle_{T}\arrowvert0\rangle_{E_{1}})\nonumber\\
&&\hskip -3mm {=}(U_{F}\otimes I_{E_{1}})\arrowvert0\rangle_{E_{2}}(\arrowvert+,e_{+,+}^{B}\rangle+\arrowvert-,e_{+,-}^{B}\rangle)_{TE_{1}}\nonumber\\
&& \hskip -3mm {=}\arrowvert e_{0,0}^{A}\rangle_{E_{2}}\arrowvert+\rangle_{T}(\arrowvert e_{+,+}^{B}\rangle+\arrowvert e_{+,-}^{B}\rangle)_{E_{1}}\nonumber\\
&&+\arrowvert e_{0,1}^{A}\rangle_{E_{2}}\arrowvert+\rangle_{T}(\arrowvert e_{+,+}^{B}\rangle+\arrowvert e_{+,-}^{B}\rangle)_{E_{1}}\nonumber\\
 &&+\arrowvert e_{1,0}^{A}\rangle_{E_{2}}\arrowvert+\rangle_{T}(\arrowvert e_{+,+}^{B}\rangle-\arrowvert e_{+,-}^{B}\rangle)_{E_{1}}\nonumber\\
 &&+\arrowvert e_{1,1}^{A}\rangle_{E_{2}}\arrowvert+\rangle_{T}(\arrowvert e_{+,+}^{B}\rangle-\arrowvert e_{+,-}^{B}\rangle)_{E_{1}}\nonumber\\
&&+\arrowvert e_{0,0}^{A}\rangle_{E_{2}}\arrowvert-\rangle_{T}(\arrowvert e_{+,+}^{B}\rangle+\arrowvert e_{+,-}^{B}\rangle)_{E_{1}}\nonumber\\
&&-\arrowvert e_{0,1}^{A}\rangle_{E_{2}}\arrowvert-\rangle_{T}(\arrowvert e_{+,+}^{B}\rangle+\arrowvert e_{+,-}^{B}\rangle)_{E_{1}}\nonumber\\
&&+\arrowvert e_{1,0}^{A}\rangle_{E_{2}}\arrowvert-\rangle_{T}(\arrowvert e_{+,+}^{B}\rangle-\arrowvert e_{+,-}^{B}\rangle)_{E_{1}}\nonumber\\
&&-\arrowvert e_{1,1}^{A}\rangle_{E_{2}}\arrowvert-\rangle_{T}(\arrowvert e_{+,+}^{B}\rangle-\arrowvert e_{+,-}^{B}\rangle)_{E_{1}},
\end{eqnarray}
\begin{eqnarray}
V&&\hskip -7mm
(\arrowvert0\rangle_{E_{2}}\arrowvert-\rangle_{T}\arrowvert0\rangle_{E_{1}})\nonumber\\
&&\hskip -3mm {=}(U_{F}\otimes I_{E_{1}})\arrowvert0\rangle_{E_{2}}(\arrowvert+,e_{-,+}^{B}\rangle+\arrowvert-,e_{-,-}^{B}\rangle)_{TE_{1}}\nonumber\\
&&\hskip -3mm {=}\arrowvert e_{0,0}^{A}\rangle_{E_{2}}\arrowvert+\rangle_{T}(\arrowvert e_{-,+}^{B}\rangle+\arrowvert e_{-,-}^{B}\rangle)_{E_{1}}\nonumber\\
&&+\arrowvert e_{0,1}^{A}\rangle_{E_{2}}\arrowvert+\rangle_{T}(\arrowvert e_{-,+}^{B}\rangle+\arrowvert e_{-,-}^{B}\rangle)_{E_{1}}\nonumber\\
  &&+\arrowvert e_{1,0}^{A}\rangle_{E_{2}}\arrowvert+\rangle_{T}(\arrowvert e_{-,+}^{B}\rangle-\arrowvert e_{-,-}^{B}\rangle)_{E_{1}}\nonumber\\
  &&+\arrowvert e_{1,1}^{A}\rangle_{E_{2}}\arrowvert+\rangle_{T}(\arrowvert e_{-,+}^{B}\rangle-\arrowvert e_{-,-}^{B}\rangle)_{E_{1}}\nonumber\\
&&+\arrowvert e_{0,0}^{A}\rangle_{E_{2}}\arrowvert-\rangle_{T}(\arrowvert e_{-,+}^{B}\rangle+\arrowvert e_{-,-}^{B}\rangle)_{E_{1}}\nonumber\\
&&-\arrowvert e_{0,1}^{A}\rangle_{E_{2}}\arrowvert-\rangle_{T}(\arrowvert e_{-,+}^{B}\rangle+\arrowvert e_{-,-}^{B}\rangle)_{E_{1}}\nonumber\\
 &&+\arrowvert e_{1,0}^{A}\rangle_{E_{2}}\arrowvert-\rangle_{T}(\arrowvert e_{-,+}^{B}\rangle-\arrowvert e_{-,-}^{B}\rangle)_{E_{1}}\nonumber\\
 &&-\arrowvert e_{1,1}^{A}\rangle_{E_{2}}\arrowvert-\rangle_{T}(\arrowvert e_{-,+}^{B}\rangle-\arrowvert e_{-,-}^{B}\rangle)_{E_{1}}.
\end{eqnarray}
Let
\begin{eqnarray}
 \arrowvert f_{0}\rangle_{E}&&\hskip -6mm {=}\arrowvert e_{0,0}^{A}\rangle_{E_{2}}(\arrowvert e_{+,+}^{B}\rangle+\arrowvert e_{+,-}^{B}\rangle)_{E_{1}}\nonumber\\
&&\hskip -3mm+\arrowvert e_{0,1}^{A}\rangle_{E_{2}}(\arrowvert e_{+,+}^{B}\rangle+\arrowvert e_{+,-}^{B}\rangle)_{E_{1}}\nonumber\\
&&\hskip -3mm+\arrowvert e_{1,0}^{A}\rangle_{E_{2}}(\arrowvert e_{+,+}^{B}\rangle{-}\arrowvert e_{+,-}^{B}\rangle)_{E_{1}}\nonumber\\
&&\hskip -3mm+\arrowvert e_{1,1}^{A}\rangle_{E_{2}}(\arrowvert e_{+,+}^{B}\rangle{-}\arrowvert e_{+,-}^{B}\rangle)_{E_{1}},\\
\arrowvert f_{1}\rangle_{E}&&\hskip -6mm {=}\arrowvert e_{0,0}^{A}\rangle_{E_{2}}(\arrowvert e_{+,+}^{B}\rangle+\arrowvert e_{+,-}^{B}\rangle)_{E_{1}}\nonumber\\
&&\hskip -3mm-\arrowvert e_{0,1}^{A}\rangle_{E_{2}}(\arrowvert e_{+,+}^{B}\rangle+\arrowvert e_{+,-}^{B}\rangle)_{E_{1}}\nonumber\\
&&\hskip -3mm+\arrowvert e_{1,0}^{A}\rangle_{E_{2}}(\arrowvert e_{+,+}^{B}\rangle-\arrowvert e_{+,-}^{B}\rangle)_{E_{1}}\nonumber\\
&&\hskip -3mm-\arrowvert e_{1,1}^{A}\rangle_{E_{2}}(\arrowvert e_{+,+}^{B}\rangle-\arrowvert e_{+,-}^{B}\rangle)_{E_{1}},
\\
\arrowvert f_{2}\rangle_{E}&&\hskip -6mm {=}\arrowvert e_{0,0}^{A}\rangle_{E_{2}}(\arrowvert e_{-,+}^{B}\rangle+\arrowvert e_{-,-}^{B}\rangle)_{E_{1}}\nonumber\\
&&\hskip -3mm+\arrowvert e_{0,1}^{A}\rangle_{E_{2}}(\arrowvert e_{-,+}^{B}\rangle+\arrowvert e_{-,-}^{B}\rangle)_{E_{1}}\nonumber\\
&&\hskip -3mm+\arrowvert e_{1,0}^{A}\rangle_{E_{2}}(\arrowvert e_{-,+}^{B}\rangle-\arrowvert e_{-,-}^{B}\rangle)_{E_{1}}\nonumber\\
&&\hskip -3mm+\arrowvert e_{1,1}^{A}\rangle_{E_{2}}(\arrowvert e_{-,+}^{B}\rangle-\arrowvert e_{-,-}^{B}\rangle)_{E_{1}},\\
\arrowvert f_{3}\rangle_{E}&&\hskip -6mm {=}\arrowvert e_{0,0}^{A}\rangle_{E_{2}}(\arrowvert e_{-,+}^{B}\rangle+\arrowvert e_{-,-}^{B}\rangle)_{E_{1}}\nonumber\\
&&\hskip -3mm-\arrowvert e_{0,1}^{A}\rangle_{E_{2}}(\arrowvert e_{-,+}^{B}\rangle+\arrowvert e_{-,-}^{B}\rangle)_{E_{1}}\nonumber\\
&&\hskip -3mm+\arrowvert e_{1,0}^{A}\rangle_{E_{2}}(\arrowvert e_{-,+}^{B}\rangle-\arrowvert e_{-,-}^{B}\rangle)_{E_{1}}\nonumber\\
&&\hskip -3mm-\arrowvert e_{1,1}^{A}\rangle_{E_{2}}(\arrowvert e_{-,+}^{B}\rangle-\arrowvert e_{-,-}^{B}\rangle)_{E_{1}}.
\end{eqnarray}
Then, ignoring the order of particles, we have
\begin{equation}\label{V+}
V(\arrowvert0\rangle_{E_{2}}\arrowvert+\rangle_{T}\arrowvert0\rangle_{E_{1}})
=\frac{1}{2}\arrowvert+\rangle_{T}\arrowvert f_{0}\rangle_{E}+\frac{1}{2}\arrowvert-\rangle_{T}\arrowvert f_{1}\rangle_{E},
\end{equation}
\begin{equation}\label{V-}
V(\arrowvert0\rangle_{E_{2}}\arrowvert-\rangle_{T}\arrowvert0\rangle_{E_{1}})
=\frac{1}{2}\arrowvert+\rangle_{T}\arrowvert f_{2}\rangle_{E}+\frac{1}{2}\arrowvert-\rangle_{T}\arrowvert f_{3}\rangle_{E},
\end{equation}
\begin{equation}
 \langle f_{0}\arrowvert f_{0}\rangle+\langle f_{1}\arrowvert f_{1}\rangle= \langle f_{2}\arrowvert f_{2}\rangle+\langle f_{3}\arrowvert  f_{3}\rangle=1,
\end{equation}
\begin{equation}
 \langle f_{0}\arrowvert f_{2}\rangle+\langle f_{1}\arrowvert f_{3}\rangle=0.
\end{equation}
From Eqs.~(\ref{V+}) and (\ref{V-}), we can get
\begin{eqnarray}
p_{+,+}^{B}=\frac{1}{4}\langle f_{0}\arrowvert f_{0}\rangle,~~~p_{+,-}^{B}=\frac{1}{4}\langle f_{1}\arrowvert f_{1}\rangle,\\
p_{-,+}^{B}=\frac{1}{4}\langle f_{2}\arrowvert f_{2}\rangle,~~~p_{-,-}^{B}=\frac{1}{4}\langle f_{3}\arrowvert f_{3}\rangle,
\end{eqnarray}
where $p_{i,j}^{B}$ is the probability of the event that Bob's measurement result state is $\arrowvert j\rangle$ if he originally sent $\arrowvert i\rangle$, $ i, j\in\{+,-\}$.

By calculating, we get
\begin{eqnarray}
p_{+,+}^{B}-p_{+,-}^{B}
&&\hskip -6mm {=}\textrm{Re}[\langle e_{0,0}^{A}\arrowvert e_{0,1}^{A}\rangle(1 {+} 2\textrm{Re}(\langle e_{+,+}^{B}\arrowvert e_{+,-}^{B}\rangle))\nonumber\\
&&\hskip -3mm {+} \langle e_{1,0}^{A}\arrowvert e_{1,1}^{A}\rangle(1 {-} 2\textrm{Re}(\langle e_{+,+}^{B}\arrowvert e_{+,-}^{B}\rangle))]\nonumber\\
&&\hskip -3mm {+} \textrm{Re}[\langle e_{0,0}^{A}\arrowvert e_{1,1}^{A}\rangle(p_{+,+} {-} p_{+,-}\nonumber\\
&&\hskip -3mm {+} 2i\textrm{Im}(\langle e_{+,-}^{B}\arrowvert e_{+,+}^{B}\rangle))]\nonumber\\
&&\hskip -3mm+\textrm{Re}[\langle e_{0,1}^{A}\arrowvert e_{1,0}^{A}\rangle(p_{+,+} {-} p_{+,-}\nonumber\\
&&\hskip -3mm + 2i\textrm{Im}(\langle e_{+,-}^{B}\arrowvert e_{+,+}^{B}\rangle))],
\end{eqnarray}
and
\begin{eqnarray}
p_{-,+}^{B}-p_{-,-}^{B}
&&\hskip -6mm {=}\textrm{Re}[\langle e_{0,0}^{A}\arrowvert e_{0,1}^{A}\rangle(1 {+} 2\textrm{Re}(\langle e_{-,+}^{B}\arrowvert e_{-,-}^{B}\rangle))\nonumber\\
&&\hskip -3mm{+}\langle e_{1,0}^{A}\arrowvert e_{1,1}^{A}\rangle(1 {-} 2\textrm{Re}(\langle e_{-,+}^{B}\arrowvert e_{-,-}^{B}\rangle))]\nonumber\\
&&\hskip -3mm{+} \textrm{Re}[\langle e_{0,0}^{A}\arrowvert e_{1,1}^{A}\rangle( p_{-,+} {-} p_{-,-}\nonumber\\
&&\hskip -3mm{+} 2i\textrm{Im}(\langle e_{-,-}^{B}\arrowvert e_{-,+}^{B}\rangle))]\nonumber\\
&&\hskip -3mm{+}\textrm{Re}[\langle e_{0,1}^{A}\arrowvert e_{1,0}^{A}\rangle( p_{-,+}{-} p_{-,-}\nonumber\\
&&\hskip -3mm+2i\textrm{Im}(\langle e_{-,-}^{B}\arrowvert e_{-,+}^{B}\rangle))].
\end{eqnarray}

By transference, we obtain
\begin{eqnarray}
\textrm{Re}&&\hskip -7mm [\langle e_{0,0}^{A}\arrowvert e_{1,1}^{A}\rangle(p_{+,+}- p_{+,-}+2i\textrm{Im}(\langle e_{+,-}^{B}\arrowvert e_{+,+}^{B}\rangle))]~~\nonumber\\
&&\hskip-7mm= p_{+,+}^{B}{-}p_{+,-}^{B}\nonumber\\
&&\hskip-6mm{-}\textrm{Re}[\langle e_{0,1}^{A}\arrowvert e_{1,0}^{A}\rangle(p_{+,+} {-} p_{+,-}{+}2i\textrm{Im}(\langle e_{+,-}^{B}\arrowvert e_{+,+}^{B}\rangle))]\nonumber\\
&&\hskip-6mm{-} \textrm{Re}[\langle e_{0,0}^{A}\arrowvert e_{0,1}^{A}\rangle(1 {+} 2\textrm{Re}(\langle e_{+,+}^{B}\arrowvert e_{+,-}^{B}\rangle))\nonumber\\
&&\hskip-6mm{+} \langle e_{1,0}^{A}\arrowvert e_{1,1}^{A}\rangle(1 {-} 2\textrm{Re}(\langle e_{+,+}^{B}\arrowvert e_{+,-}^{B}\rangle))]
,\label{p+++-}
\end{eqnarray}
and
\begin{eqnarray}
-\textrm{Re}&&\hskip -7mm [\langle e_{0,0}^{A}\arrowvert e_{1,1}^{A}\rangle( p_{-,+}{-} p_{-,-}{+}2i\textrm{Im}(\langle e_{-,-}^{B}\arrowvert e_{-,+}^{B}\rangle))]\nonumber
\\
&&\hskip-11mm= p_{-,-}^{B}{-}p_{-,+}^{B}\nonumber
\\
&&\hskip-9mm {-}\textrm{Re}[\langle e_{0,1}^{A}\arrowvert e_{1,0}^{A}\rangle( p_{-,-} {-} p_{-,+}){+}2i\textrm{Im}(\langle e_{-,-}^{B}\arrowvert e_{-,+}^{B}\rangle)] \nonumber
\\
&&\hskip-9mm {-} \textrm{Re}[\langle e_{0,0}^{A}\arrowvert e_{0,1}^{A}\rangle(-1-2\textrm{Re}(\langle e_{-,+}^{B}\arrowvert e_{-,-}^{B}\rangle))\nonumber
\\
&&\hskip-9mm {+}\langle e_{1,0}^{A}\arrowvert e_{1,1}^{A}\rangle(2\textrm{Re}(\langle e_{-,+}^{B}\arrowvert e_{-,-}^{B}\rangle){-}1)].\label{p---+}
\end{eqnarray}
Adding the corresponding two sides of Eq.~(\ref{p+++-}) and Eq.~(\ref{p---+}),
\begin{eqnarray}\label{leftandright}
\textrm{Re}&&\hskip-6mm[\langle e_{0,0}^{A}\arrowvert e_{1,1}^{A}\rangle(p_{+,+} {-} p_{+,-} {+} 2i\textrm{Im}(\langle e_{+,-}^{B}\arrowvert e_{+,+}^{B}\rangle))]\nonumber
-\\
&&\hskip-2mm\textrm{Re}[\langle e_{0,0}^{A}\arrowvert e_{1,1}^{A}\rangle( p_{-,+} {-} p_{-,-} {+} 2i\textrm{Im}(\langle e_{-,-}^{B}\arrowvert e_{-,+}^{B}\rangle))]\nonumber
\\
&&\hskip-6mm{=} p_{+,+}^{B}-p_{+,-}^{B}+p_{-,-}^{B}-p_{-,+}^{B}\nonumber\\
&&\-\textrm{Re}[\langle e_{0,0}^{A}\arrowvert e_{0,1}^{A}\rangle(2\textrm{Re}(\langle e_{+,+}^{B}\arrowvert e_{+,-}^{B}\rangle)\nonumber\\
&&-2\textrm{Re}(\langle e_{-,+}^{B}\arrowvert e_{-,-}^{B}\rangle))]\nonumber
\\
&&-\textrm{Re}[\langle e_{1,0}^{A}\arrowvert e_{1,1}^{A}\rangle(2\textrm{Re}\langle e_{-,+}^{B}\arrowvert e_{-,-}^{B}\rangle\nonumber\\
&&-2\textrm{Re}\langle e_{+,+}^{B}\arrowvert e_{+,-}^{B}\rangle)]\nonumber
\\
&&-\textrm{Re}[\langle e_{0,1}^{A}\arrowvert e_{1,0}^{A}\rangle(p_{+,+}{-} p_{+,-}\nonumber\\
&&+2i\textrm{Im}(\langle e_{+,-}^{B}\arrowvert e_{+,+}^{B}\rangle))]\nonumber\\
&&-\textrm{Re}[\langle e_{0,1}^{A}\arrowvert e_{1,0}^{A}\rangle( p_{-,-} {-} p_{-,+} \nonumber\\
&&+ 2i\textrm{Im}(\langle e_{-,-}^{B}\arrowvert e_{-,+}^{B}\rangle))]\\
&&\hskip-6mm{=} 2{-}2(p_{+,-}^{B}{+}p_{-,+}^{B})\nonumber
\\
&&-\textrm{Re}[\langle e_{0,0}^{A}\arrowvert e_{0,1}^{A}\rangle(2\textrm{Re}(\langle e_{+,+}^{B}\arrowvert e_{+,-}^{B}\rangle) \nonumber\\
&&- 2\textrm{Re}(\langle e_{-,+}^{B}\arrowvert e_{-,-}^{B}\rangle))]\nonumber
\\
&& -\textrm{Re}[\langle e_{1,0}^{A}\arrowvert e_{1,1}^{A}\rangle(2\textrm{Re}\langle e_{-,+}^{B}\arrowvert e_{-,-}^{B}\rangle\nonumber\\
&&-2\textrm{Re}\langle e_{+,+}^{B}\arrowvert e_{+,-}^{B}\rangle)]\nonumber
\\
&&-\textrm{Re}[\langle e_{0,1}^{A}\arrowvert e_{1,0}^{A}\rangle(p_{+,+}{-} p_{+,-} \nonumber\\
&&+ 2i\textrm{Im}(\langle e_{+,-}^{B}\arrowvert e_{+,+}^{B}\rangle))]\nonumber
\\
&& -\textrm{Re}[\langle e_{0,1}^{A}\arrowvert e_{1,0}^{A}\rangle( p_{-,-}{-} p_{-,+} \nonumber\\
&&+ 2i\textrm{Im}(\langle e_{-,-}^{B}\arrowvert e_{-,+}^{B}\rangle))].
\end{eqnarray}
Observe that, for any two vectors $\arrowvert x_{1}\rangle$ and $\arrowvert x_{2}\rangle$,
\begin{eqnarray}
\arrowvert \textrm{Re}(\langle x_{1}\arrowvert x_{2}\rangle) \arrowvert \leq \arrowvert\langle x_{1}\arrowvert x_{2}\rangle\arrowvert \leq \sqrt{\langle x_{1}\arrowvert x_{1}\rangle\langle x_{2}\arrowvert x_{2}\rangle}~
\end{eqnarray}
 and
\begin{eqnarray}
\arrowvert \textrm{Im}(\langle x_{1}\arrowvert x_{2}\rangle) \arrowvert \leq \arrowvert\langle x_{1}\arrowvert x_{2}\rangle\arrowvert \leq \sqrt{\langle x_{1}\arrowvert x_{1}\rangle\langle x_{2}\arrowvert x_{2}\rangle},~
\end{eqnarray}
%i.e.
% \begin{eqnarray}
% \textrm{Re}(\langle x_{1}\arrowvert x_{2}\rangle) \in [-\sqrt{\langle x_{1}\arrowvert x_{1}\rangle\langle x_{2}\arrowvert x_{2}\rangle},\sqrt{\langle x_{1}\arrowvert x_{1}\rangle\langle x_{2}\arrowvert x_{2}\rangle}]~~~~~~
%\end{eqnarray}
% and
% \begin{eqnarray}
% \textrm{Im}(\langle x_{1}\arrowvert x_{2}\rangle) \in [-\sqrt{\langle x_{1}\arrowvert x_{1}\rangle\langle x_{2}\arrowvert x_{2}\rangle},\sqrt{\langle x_{1}\arrowvert x_{1}\rangle\langle x_{2}\arrowvert x_{2}\rangle}]\!.~~~~~~
% \end{eqnarray}
 Thus,
\begin{eqnarray}\label{left1}
\textrm{Re}&&\hskip-7mm[\langle e_{0,0}^{A}\arrowvert e_{1,1}^{A}\rangle(p_{+,+}- p_{+,-}+2i\textrm{Im}(\langle e_{+,-}^{B}\arrowvert e_{+,+}^{B}\rangle))]\nonumber\\
&&\hskip-8mm{=}\textrm{Re}[\langle e_{0,0}^{A}\arrowvert e_{1,1}^{A}\rangle(1{-}2 p_{+,-}{+}2i\textrm{Im}(\langle e_{+,-}^{B}\arrowvert e_{+,+}^{B}\rangle))]\nonumber\\
&&\hskip-8mm{\leq}\arrowvert \langle e_{0,0}^{A}\arrowvert e_{1,1}^{A}\rangle \arrowvert \arrowvert 1{-}2 p_{+,-} {+}2i\textrm{Im}(\langle e_{+,-}^{B}\arrowvert e_{+,+}^{B}\rangle) \arrowvert\nonumber\\
&&\hskip-8mm{=} \arrowvert \langle e_{0,0}^{A}\arrowvert e_{1,1}^{A}\rangle \arrowvert \sqrt{ (1{-}2 p_{+,-})^{2} {+}4\textrm{Im}^{2}(\langle e_{+,-}^{B}\arrowvert e_{+,+}^{B}\rangle)}\nonumber\\
&&\hskip-8mm{\leq} \arrowvert \langle e_{0,0}^{A}\arrowvert e_{1,1}^{A}\rangle \arrowvert \sqrt{ (1{-}2 p_{+,-})^{2} {+}4 p_{+,-} p_{+,+})}\nonumber\\
&&\hskip-8mm{=} \arrowvert \langle e_{0,0}^{A}\arrowvert e_{1,1}^{A}\rangle \arrowvert \sqrt{ 1{-}4 p_{+,-}{+}4 p^{2}_{+,-}{+}4 p_{+,-}(1{-} p_{+,-})}\nonumber\\
&&\hskip-8mm{=} \arrowvert \langle e_{0,0}^{A}\arrowvert e_{1,1}^{A}\rangle \arrowvert.
\end{eqnarray}
Similarly, we have
\begin{eqnarray}\label{left2}
-\textrm{Re}&&\hskip-7mm[\langle e_{0,0}^{A}\arrowvert e_{1,1}^{A}\rangle( p_{-,+}- p_{-,-}{+}2i\textrm{Im}(\langle e_{-,-}^{B}\arrowvert e_{-,+}^{B}\rangle))]\nonumber\\
&&\hskip-8mm{\leq}\arrowvert \langle e_{0,0}^{A}\arrowvert e_{1,1}^{A}\rangle \arrowvert,
\end{eqnarray}
\begin{eqnarray}\label{right1}
-\textrm{Re}&&\hskip-7mm[\langle e_{0,1}^{A}\arrowvert e_{1,0}^{A}\rangle(p_{+,+}- p_{+,-} {+} 2i\textrm{Im}(\langle e_{+,-}^{B}\arrowvert e_{+,+}^{B}\rangle))]\nonumber\\
&&\hskip-8mm{\leq}\arrowvert \langle e_{0,1}^{A}\arrowvert e_{1,0}^{A}\rangle \arrowvert,
\end{eqnarray}
\begin{eqnarray}\label{right2}
-\textrm{Re}&&\hskip -7mm [\langle e_{0,1}^{A}\arrowvert e_{1,0}^{A}\rangle( p_{-,-}{-} p_{-,+} {+} 2i\textrm{Im}(\langle e_{+,-}^{B}\arrowvert e_{-,-}^{B}\rangle))]\nonumber\\
&&\hskip-8mm{\leq}\arrowvert \langle e_{0,1}^{A}\arrowvert e_{1,0}^{A}\rangle \arrowvert.
\end{eqnarray}
Note that,  for any two vectors $\arrowvert x\rangle$, $\arrowvert y\rangle$, it holds$\arrowvert x-y\arrowvert \leq \arrowvert x\arrowvert + \arrowvert y\arrowvert$, $\arrowvert x+y\arrowvert \leq \arrowvert x\arrowvert + \arrowvert y\arrowvert$. Using these facts, we get
\begin{eqnarray}\label{right3}
-\textrm{Re}&&\hskip -7mm[\langle e_{0,0}^{A}\arrowvert e_{0,1}^{A}\rangle(2\textrm{Re}(\langle e_{+,+}^{B}\arrowvert e_{+,-}^{B}\rangle) \nonumber\\
&&  -2\textrm{Re}(\langle e_{-,+}^{B}\arrowvert e_{-,-}^{B}\rangle)]\nonumber
\\
&&\hskip-8mm{\leq}\arrowvert\langle e_{0,0}^{A}\arrowvert e_{0,1}^{A}\rangle(2\textrm{Re}(\langle e_{+,+}^{B}\arrowvert e_{+,-}^{B}\rangle)\nonumber\\
&&-2\textrm{Re}(\langle e_{-,+}^{B}\arrowvert e_{-,-}^{B}\rangle)\arrowvert\nonumber
\\
&&\hskip-8mm{\leq} \arrowvert\langle e_{0,0}^{A}\arrowvert e_{0,1}^{A}\rangle\arrowvert(\arrowvert2\textrm{Re}(\langle e_{+,+}^{B}\arrowvert e_{+,-}^{B}\rangle) \nonumber\\
&&- 2\textrm{Re}(\langle e_{-,+}^{B}\arrowvert e_{-,-}^{B}\rangle)\arrowvert)\nonumber
\\
&&\hskip-8mm{\leq} \arrowvert\langle e_{0,0}^{A}\arrowvert e_{0,1}^{A}\rangle\arrowvert(\arrowvert2\textrm{Re}(\langle e_{+,+}^{B}\arrowvert e_{+,-}^{B}\rangle)\arrowvert\nonumber\\
&&+\arrowvert2\textrm{Re}(\langle e_{-,+}^{B}\arrowvert e_{-,-}^{B}\rangle)\arrowvert)\nonumber
\\
&&\hskip-8mm{\leq} 2\arrowvert\langle e_{0,0}^{A}\arrowvert e_{0,1}^{A}\rangle\arrowvert(2\arrowvert\langle e_{+,+}^{B}\arrowvert e_{+,-}^{B}\rangle\arrowvert \nonumber\\
&&+2\arrowvert\langle e_{-,+}^{B}\arrowvert e_{-,-}^{B}\rangle\arrowvert).
\end{eqnarray}
%第二个不等式做了修改
Similarly, we have
 \begin{eqnarray}\label{right4}
\textrm{Re}&&\hskip -7mm[\langle e_{1,0}^{A}\arrowvert e_{1,1}^{A}\rangle(2\textrm{Re}(\langle e_{-,+}^{B}\arrowvert e_{-,-}^{B}\rangle)\nonumber\\
&& -2 \textrm{Re}(\langle e_{+,+}^{B}\arrowvert e_{+,-}^{B}\rangle)]\nonumber
\\
&&\hskip-7mm{\leq} 2\arrowvert\langle e_{1,0}^{A}\arrowvert e_{1,1}^{A}\rangle\arrowvert(2\arrowvert\langle e_{-,+}^{B}\arrowvert e_{-,-}^{B}\rangle\arrowvert\nonumber\\
&&+2\arrowvert\langle e_{+,+}^{B}\arrowvert e_{+,-}^{B}\rangle\arrowvert).
\end{eqnarray}
Using $\arrowvert\langle x_{1}\arrowvert x_{2}\rangle\arrowvert{\leq} \sqrt{\langle x_{1}\arrowvert x_{1}\rangle\langle x_{2}\arrowvert x_{2}\rangle}$ and Eqs.~(\ref{leftandright}) and (\ref{left1})--(\ref{right4}), according to $\langle e_{i,j}^{A}\arrowvert e_{i,j}^{A}\rangle=p_{i,j}^{A}$ and $\langle e_{i,j}^{B}\arrowvert e_{i,j}^{B}\rangle=p_{i,j}^{B}$, we gain
 \begin{eqnarray}
&&\hskip-7mm2\arrowvert \langle e_{0,0}^{A}\arrowvert e_{1,1}^{A}\rangle \arrowvert \nonumber\\
&&\hskip-5mm\geq \textrm{Re}[\langle e_{0,0}^{A}\arrowvert e_{1,1}^{A}\rangle(p_{+,+} {-} p_{+,-} {+} 2i\textrm{Im}(\langle e_{+,-}^{B}\arrowvert e_{+,+}^{B}\rangle))]\nonumber\\
&&\hskip-4mm{-}\textrm{Re}[\langle e_{0,0}^{A}\arrowvert e_{1,1}^{A}\rangle( p_{-,+}{-} p_{-,-}{+}2i\textrm{Im}(\langle e_{-,-}^{B}\arrowvert e_{-,+}^{B}\rangle))]\nonumber\\
&&\hskip-5mm\geq2{-}2(p_{+,-}^{B}{+}p_{-,+}^{B}){-}2\arrowvert\langle e_{0,1}^{A}\arrowvert e_{1,0}^{A}\rangle \arrowvert\nonumber
\\
&&\hskip-4mm{-}2\arrowvert\langle e_{0,0}^{A}\arrowvert e_{0,1}^{A}\rangle\arrowvert(2\arrowvert\langle e_{+,+}^{B}\arrowvert e_{+,-}^{B}\rangle\arrowvert {+}2\arrowvert\langle e_{-,+}^{B}\arrowvert e_{-,-}^{B}\rangle\arrowvert)\nonumber
\\
&&\hskip-4mm{-}2\arrowvert\langle e_{1,0}^{A}\arrowvert
e_{1,1}^{A}\rangle\arrowvert(2\arrowvert\langle e_{-,+}^{B}\arrowvert e_{-,-}^{B}\rangle\arrowvert{+}2\arrowvert\langle e_{+,+}^{B}\arrowvert e_{+,-}^{B}\rangle\arrowvert)\nonumber
\\
&&\hskip-5mm\geq2{-}2(p_{+,-}^{B}{+}p_{-,+}^{B})\nonumber
\\
&&\hskip-4mm{-}4\sqrt{p^{A}_{0,0}p^{A}_{0,1}}(\sqrt{p^{B}_{+,+}p^{B}_{+,-}} {+}\sqrt{p^{B}_{-,+}p^{B}_{-,-}})\nonumber
\\
&&\hskip-4mm{-}4\sqrt{p^{A}_{1,0}p^{A}_{1,1}}(\sqrt{p^{B}_{-,+}p^{B}_{-,-}} {+}\sqrt{p^{B}_{+,+}p^{B}_{+,-}}).
 \end{eqnarray}

Furthermore, we can learn
\begin{eqnarray}
\arrowvert \langle e_{0,0}^{A}\arrowvert e_{1,1}^{A}\rangle \arrowvert &&\hskip-7mm{\geq} 1{-}(p_{+,-}^{B}{+}p_{-,+}^{B})\nonumber\\
&&\hskip-3mm{-}2\sqrt{p^{A}_{0,0}p^{A}_{0,1}}(\sqrt{p^{B}_{+,+}p^{B}_{+,-}}{+}\sqrt{p^{B}_{-,+}p^{B}_{-,-}})\nonumber\\
&&\hskip-3mm{-}2\sqrt{p^{A}_{1,0}p^{A}_{1,1}}(\sqrt{p^{B}_{-,+}p^{B}_{-,-}}{+}\sqrt{p^{B}_{+,+}p^{B}_{+,-}})\nonumber\\
&&\hskip-7mm {=}C.
\end{eqnarray}

It's clear that $p_{+,+}{\leq}1$ and $ p_{-,-}{\leq}1$. Further we have $\arrowvert\langle e_{+,+}^{B}\arrowvert e_{-,-}^{B}\rangle\arrowvert{\leq}1$. %Assuming $C$ is non-negative, it may be used to lower bound $\arrowvert\langle e_{0,0}^{A}\arrowvert e_{1,1}^{A}\rangle\arrowvert^{2}$.
Let
\begin{equation}
\widetilde{C}=\left\{
\begin{aligned}
C^{2},~~~\textrm{if}~C\geq0,\\
0,~~~\textrm{otherwise},
\end{aligned}
\right.
\end{equation}
then $\arrowvert\langle e_{0,0}^{A}\arrowvert e_{1,1}^{A}\rangle\arrowvert^{2}\geq \widetilde{C}$. Since $\arrowvert\langle e_{0,0}^{A}\arrowvert e_{1,1}^{A}\rangle\arrowvert^{2}$ being always non-negative, it makes sense to let $\widetilde{C}$ equal to 0 if $C<0$. Thus, $S(\widetilde{\sigma}_{1})=h(\widetilde{\lambda}_{+})\leq h(\widetilde{\lambda})$, where
\begin{equation}
\widetilde{\lambda}=\frac{1}{2}+\frac{\sqrt{(p_{0,0}^{A}-p_{1,1}^{A})^{2}+4\widetilde{C}}}{2(p_{0,0}^{A}+p_{1,1}^{A})}.
\end{equation}
\subsection{Final Key Rate Bound}\label{subsec2}
From the all above discussion, we know that the key rate is lower-bounded by
\begin{eqnarray}\label{eq8}
r&&\hskip-7mm{\geq} H(\frac{1}{2}p_{0,0}^{A},\frac{1}{2}p_{0,1}^{A},\frac{1}{2}p_{1,0}^{A},\frac{1}{2}p_{1,1}^{A})\nonumber\\
&&\hskip-5mm-H(\frac{1}{2}(p_{0,0}^{A}+p_{1,1}^{A}),\frac{1}{2}(p_{0,1}^{A}+p_{1,0}^{A}))\nonumber\\
&&\hskip-5mm -\frac{1}{2}(p_{0,1}^{A},P_{1,0}^{A})-\frac{1}{2}(p_{0,0}^{A},P_{1,1}^{A})h(\widetilde{\lambda})+1\nonumber\\
&&\hskip-5mm-H(\frac{1}{2}p_{0,0}^{A},\frac{1}{2}p_{0,1}^{A},\frac{1}{2}p_{1,0}^{A},\frac{1}{2}p_{1,1}^{A})\nonumber\\
&&\hskip-7mm{=} 1-H(\frac{1}{2}(p_{0,0}^{A}+p_{1,1}^{A}),\frac{1}{2}(p_{0,1}^{A}{+}p_{1,0}^{A}))\nonumber\\
&&\hskip-5mm-\frac{1}{2}(p_{0,1}^{A}+P_{1,0}^{A})-\frac{1}{2}(p_{0,0}^{A}+P_{1,1}^{A})h(\widetilde{\lambda}).~
\end{eqnarray}
Let $\widetilde{r}=1-H(\frac{1}{2}(p_{0,0}^{A}+p_{1,1}^{A}),\frac{1}{2}(p_{0,1}^{A}+p_{1,0}^{A}))-\frac{1}{2}(p_{0,1}^{A}+P_{1,0}^{A})$
$-\frac{1}{2}(p_{0,0}^{A}+P_{1,1}^{A})h(\widetilde{\lambda})$.
We can compute the lower bound $\widetilde{r}$ according to the parameters that Alice and Bob may be estimated.

\subsection{Examples}\label{subsec2}
Let's now demonstrate our key rate bound on certain examples. In particular, we assume that Eve's attack is symmetric in that it can be characterized as follows

1. Let $Q_{Z}$ be error rate on the $Z$ basis.
Then, $Q_{Z}=p_{0,1}^{A}=p_{1,0}^{A}$.

2. Let $Q_{X}$ be error rate on the $X$ basis.
Then, $Q_{X}=p_{+,-}^{B}=p_{-,+}^{B}$.

Alice and Bob may estimate these two parameters, we can enforce the restriction that Eve use such a symmetric attack, thus, we have

\begin{eqnarray}
p_{0,1}^{A}{=}p_{1,0}^{A}{=}Q_{Z},~p_{0,0}^{A}{=}p_{1,1}^{A}{=}1-Q_{Z},
\end{eqnarray}
\begin{eqnarray}
p_{+,-}^{B}{=}p_{+,-}^{B}{=}Q_{X},~p_{+,+}^{B}{=}p_{-,-}^{B}{=}1-Q_{X}\!.~
\end{eqnarray}

Indeed, as long as $r>0$, Alice and Bob may distill a secret key. In order to understand the effects of $Q_{Z}$ and $Q_{X}$ on $\widetilde{r}$, three cases, $Q_{Z}=\frac{Q}{2}$, $Q_{Z}=Q$, and $Q_{Z}=2Q$, are considered.
Furthermore, for each case, we consider three subcases, $Q_{X}=\frac{Q}{2}$, $Q_{X}=Q$, and $Q_{X}=2Q$.
For $\widetilde{r}\geq 0$, the allowed maximum values of $Q$ are summarized in Table~\ref{tab1}.
\begin{table}
    \center
    \caption{The maximum values of $Q$ for $\widetilde{r}>0$.}
    \label{tab1}
   \renewcommand\arraystretch{1.3}%\resizebox{10cm}{1.5cm}{
    \begin{tabular}{cccc}
        \hline
        ~&$Q_{X}=\frac{Q}{2}$&$Q_{X}=Q$ &$Q_{X}=2Q$ \\
        \hline
        $Q_{Z}=\frac{Q}{2}$ & $8.91\%$ & $6.57\%$ & $4.71\%$ \\
        $Q_{Z}=Q$ & $5.89\%$ & $4.46\%$ &$3.29\%$ \\
        $~Q_{Z}=2Q~~$ & $~~4.42\%~~$ & $~~3.34\%~~$ & $~~2.49\%~$ \\
        \hline
    \end{tabular}%}
\end{table}

Since $r\geq\widetilde{r}\geq0$, we can obtain the following results.

\emph{Case~1:} $Q_{Z}=\frac{Q}{2}$. By the values of $Q_{X}$, three subcases are as follows:

~~~~$\cdot$ If $Q_{X}=\frac{Q}{2}$, then $r>0$ for $Q<8.91\%$;

~~~~$\cdot$ If $Q_{X}=Q$, then $r>0$ for $Q<6.57\%$;

\begin{figure}%[hdtp]
  \centering
  \includegraphics[width=0.48\textwidth]{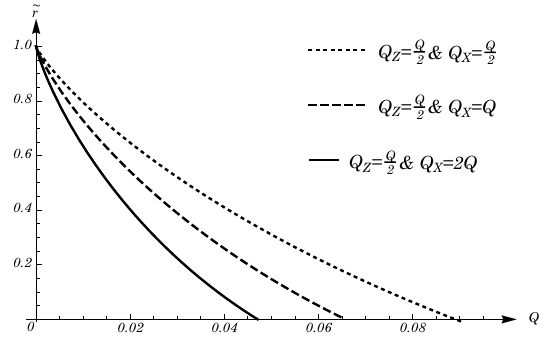}\\
  \caption{
  Key rate bound $\widetilde{r}$ as $Q_{Z}=\frac{Q}{2}$.
  }
  \label{fig1}
\end{figure}

~~~~$\cdot$ If $Q_{X}=2Q$, then $r>0$ for $Q<4.71\%$.
\newline
In Case 1,
the graphs of the key rate bound, $\widetilde{r}$, when $Q_{X}=\frac{Q}{2}$, $Q_{X}=Q$, and $Q_{X}=2Q$, % as a function of $Q$,
are seen in Fig. \ref{fig1}.

\emph{Case~2:} $Q_{Z}=Q$. By the values of $Q_{X}$, three subcases are as follows:

~~~~$\cdot$ If $Q_{X}=\frac{Q}{2}$, then $r>0$ for $Q<5.89\%$;

~~~~$\cdot$ If $Q_{X}=Q$, then $r>0$ for $Q<4.46\%$;

\begin{figure}[!h]
\centering
  \includegraphics[width=0.48\textwidth]{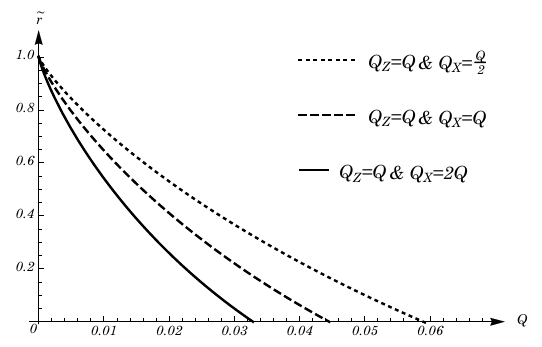}\\
  \caption{ Key rate bound $\widetilde{r}$ as $Q_{Z}=Q$.}
  \label{fig2}
\end{figure}

~~~~$\cdot$ If $Q_{X}=2Q$, then $r>0$ for $Q<3.29\%$.
\newline
In Case 2,
the graphs of the key rate bound, $\widetilde{r}$, when $Q_{X}=\frac{Q}{2}$, $Q_{X}=Q$, and $Q_{X}=2Q$, % as a function of $Q$,
are seen in Fig. \ref{fig2}.

\emph{Case~3:} $Q_{Z}=2Q$. By the values of $Q_{X}$, three subcases are as follows:

~~~~$\cdot$ If $Q_{X}=\frac{Q}{2}$, then $r>0$ for $Q<4.42\%$;

~~~~$\cdot$ If $Q_{X}=Q$, then $r>0$ for $Q<3.34\%$;

\begin{figure}%[hdtp]
  \centering
  \includegraphics[width=0.48\textwidth]{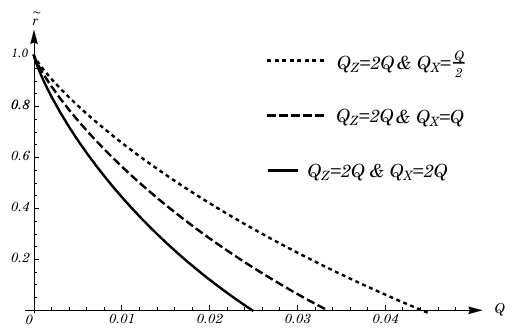}\\
  \caption{ Key rate bound $\widetilde{r}$ as $Q_{Z}=2Q$.}
  \label{fig3}
\end{figure}

~~~~$\cdot$ If $Q_{X}=2Q$, then $r>0$ for $Q<2.49\%$.
\newline
In Case 3, the graphs of the key rate bound, $\widetilde{r}$, when $Q_{X}=\frac{Q}{2}$, $Q_{X}=Q$, and $Q_{X}=2Q$, % as a function of $Q$,
are seen in Fig. \ref{fig3}.

\subsection{Security against joint attacks}\label{subsec2}
We have proven that the protocol constructed is unconditionally secure against collective attacks.
Note that, after the protocol, Bob and Alice may symmetrize their raw key by using a randomly chosen and publicly disclosed permutation.
This makes the protocol permutation invariant.
In this case, as shown in Refs.~\cite{renner2007symmetry} and~\cite{christandl2009postselection}, the security against collective attacks is sufficient to prove the security against any arbitrary general attack.
Thus, this protocol's unconditional security have been proved by the above.
%From the above results, it is clear that one of the important factors to this SQKD protocol's key rate, is the noise in the SIFT qubits. This makes sense, as any noise in the CTRL qubits don't directly lead to an error in Alice and Bob's raw key bit. Thus, they don't lead to additional information leaking due to error correction.

\section{Conclusion}\label{sec5}
From the perspective of resource theory, it is interesting to achieve the same quantum task using as few quantum resources as possible.
In this paper, we constructed the first SQKD protocol which restricts the quantum Bob to prepare quantum states in only one basis and removes the classical Alice's measurement capability.
Furthermore, we derived a lower bound on the key rate of the constructed protocol in the asymptotic scenario.
This indicates that the constructed protocol is unconditionally secure.

%\bibliographystyle{quantum}
%\bibliography{ckwx}

\onecolumn
\appendix

\end{document}